\documentclass[12pt,tightenlines,eqsecnum,floats,nofootinbib,aps,prd,longbibliography]{revtex4-2} 
\usepackage[colorlinks,citecolor=blue,linkcolor=black,urlcolor=blue]{hyperref}
\usepackage{enumitem}
\usepackage{graphicx,verbatim}
\usepackage{amsmath}
\usepackage{amsfonts}
\usepackage{amssymb}
\usepackage{epstopdf}
\usepackage[utf8]{inputenc}
\usepackage[strings]{underscore} 

\def\t{\tilde}
\def\h{\hat}
\def\b{\bar}
\def\f{\frac}
\def\Im{{\rm Im}}
\def\Re{{\rm Re}}
\def\rmd{\mathrm{d}}
\def\l{\ell}

\def\Lie{\mathcal{L}}
\def\scri{\mathfrak{I}}
\def\scrip{\scri^{+}}
\def\scrim{\scri^{-}}

\def\4z{\Psi_{4}^{\circ}}
\def\3z{\Psi_{3}^{\circ}}
\def\2z{\Psi_{2}^{\circ}}
\def\1z{\Psi_{1}^{\circ}}
\def\0z{\Psi_{0}^{\circ}}
\def\sigmaz{\sigma^{\circ}}
\def\sigmabz{\bar{\sigma}^{\circ}}
\def\iz{i^{\circ}}
\def\hpz{h_+^{\circ}}
\def\hcz{h_{\times}^{\circ}}

\def\qo{\mathring{q}}
\def\Do{\mathring{D}}
\def\no{\mathring{n}}
\def\lo{\mathring{\ell}}
\def\mo{\mathring{m}}
\def\mbo{\mathring{\b{m}}}
\def\d2vo{\rmd^{2} \mathring{V}}

\def\S{\mathcal{S}}
\def\Lor{\mathfrak{L}}
\def\B{\mathfrak{B}} 
\def\P{\mathcal{P}}
\def\T{\mathcal{T}}
\def\F{\mathcal{F}}

\def\be{\begin{equation}}
\def\ee{\end{equation}}
\def\ba{\begin{eqnarray}}
\def\ea{\end{eqnarray}}

\begin{document}

\title{Compact binary coalescences: Constraints on waveforms}
 
\author{Abhay Ashtekar}
\email{ashtekar.gravity@gmail.com} 
\author{Tommaso De Lorenzo}
\email{tdelorenzo@psu.edu} 
\author{Neev Khera}
\email{neevkhera@psu.edu}
\affiliation{Institute for Gravitation and the
Cosmos \& Physics Department, Penn State, University Park, PA 16802,
U.S.A.}

\begin{abstract}
  Gravitational waveforms for compact binary coalescences (CBCs) have been invaluable for detections by the LIGO-Virgo collaboration. They are obtained by a combination of semi-analytical models and numerical simulations. So far systematic errors arising from these procedures appear to be less than statistical ones. However, the significantly enhanced sensitivity of the new detectors that will become operational in the near future will require waveforms to be much more accurate.  This task would be facilitated if one has a variety of cross-checks  to \emph{evaluate} accuracy, particularly in the regions of parameter space where numerical simulations are sparse. Currently errors are estimated by comparing the candidate waveforms with the numerical relativity (NR) ones, which are taken to be exact. The goal of this paper is to propose a qualitatively different tool. We show that full non-linear general relativity (GR) imposes an infinite number of sharp constraints on the CBC waveforms. These can provide clear-cut measures to evaluate the accuracy of candidate waveforms against exact GR, help find systematic errors, and also provide external checks on NR simulations themselves. 
\end{abstract}

\maketitle

\section{Introduction}
\label{s1}
Accurate gravitational waveform models for Compact Binary Coalescences (CBCs) are crucial in Gravitational Wave (GW) data analysis. Matched-filtering used for detection, as well as procedures for parameters estimation and tests of General Relativity (GR), all rely on the precise knowledge of GW signals as predicted by GR. Since the exact two-body problem in GR is not solvable analytically, current available waveforms are produced by astute combinations of Numerical Relativity (NR) simulations (see, e.g., \cite{Boyle:2019kee})  and analytical approximations (see, e.g., \cite{phenom,PhysRevD.82.064016,PhysRevD.93.044006,PhysRevD.93.044007,PhysRevLett.113.151101,Damour2016,eob,PhysRevD.89.061502,Bohe:2016gbl,PhysRevD.89.084006,nagar2019PRD:99.044007,nagar2018PRD:98.104052}),  or interpolation methods (see, e.g., \cite{nr-surrogates,SurrHyb,Rifat:2019ltp,Lackey:2018zvw}). Thanks to the remarkable accuracy reached, one can argue that the systematic errors are 
less than statistical ones for current detectors \cite{Kumar:2016dhh,Purrer:2019jcp}. However,  since the sensitivity of current detectors is about to increase \cite{TheLIGOScientific:2014jea,TheVirgo:2014hva}, and new ground \cite{Aso:2013eba,Unnikrishnan:2013qwa,Punturo:2010zz,Reitze:2019iox} and space \cite{Audley:2017drz,Luo:2015ght,Kawamura:2006up} based detectors will soon become operational, the statistical uncertainties will reduce considerably. This in turn requires improvement in waveform models to match this enhanced accuracy \cite{Purrer:2019jcp,ab}. This is particularly important when observations are used as means of test of GR: departure from GR may remain hidden, or may be erroneously claimed due to inaccuracy of waveforms \cite{Yunes:2009ke}. Considerable ongoing effort is aimed at further reducing systematic errors in the waveform models, for instance by considering higher modes \cite{PhenomHM,Cotesta:2018fcv,PhenomPv3HM}, by including precession more carefully \cite{Khan:2018fmp,Blackman:2017dfb,Apostolatos:1994mx,Varma:2019csw}, or by improving numerical code efficiency and precision \cite{Kidder:2016hev}.   
 
Improvements of waveforms require reliable procedures to measure and evaluate their accuracy \emph{relative to the predictions of exact GR}. However, there is an immediate obstacle: the \emph{exact} waveform is not known! The current extensive and successful work typically estimates errors (see e.g. \cite{Pan:2007nw,Lindblom:2008cm,mp-followup,Kumar:2015tha,Kumar:2016dhh,Purrer:2019jcp}) by considering NR simulations as practical substitutes of the exact waveform. The reliability of such procedures strongly depends on how well one can identify sources of errors in the numerical simulations, and estimate their magnitude. This task is typically carried out using internal checks, comparing different simulations that use, e.g., different grid resolutions and extraction radii for the same physical system. While several modes of NR waveforms have been shown to pass these convergence tests,  this is not yet the case for all modes of interest.

In this paper we propose a qualitatively different tool to test the accuracy of semi-analytical models that does not rely on NR, and can therefore be used as an external check on numerical simulations as well. Our strategy is to use asymptotic symmetries ---the Bondi, Metzner, Sachs (BMS) group \cite{bondi,sachs1,bondi-sachs}--- in conjunction with the boundary conditions in the distant past and distant future normally assumed in CBCs.  Together, they provide an infinite set of constraints on the CBC waveforms. Violations of these sharp constraints by any proposed waveform provide a measure of its deviation from the exact GR prediction, \emph{without knowing} what that prediction is. Thus, it is a novel approach that can complement the current NR-based procedures.

The paper is organized as follows. In Sect.~\ref{s2} we obtain the infinite number of constraints on the gravitational waveforms. In this discussion, we have made a special effort to 
communicate the relevant results from mathematical relativity to the waveform community. 
Section \ref{s3} summarizes the main results and presents a few examples of work in progress to illustrate how they could be used to further improve our understanding of waveforms. Indeed, these results have already had some applications:\\
 (i) Using gravitational memory as an inferred observable, posterior distributions have been calculated for LIGO events reported in the first Gravitational Wave Transient Catalog (GWTC-1), using the Phenomenological (Phenom) and Effective One Body (EOB) waveforms \cite{adlkk}.  It is shown that comparison between these distributions for various angular modes of total memory can serve as diagnostic tools to further improve the waveforms. \\
 (ii)  The balance laws discussed in Sect.~\ref{s2.3.3} have been used to correct the strain waveforms in the SXS catalog to include the (evolution of the) memory that was missing in the earlier waveforms \cite{mikbdLkmpst}.\\
 (iii) The supertranslation ambiguity in the difference between the initial and the final total angular momentum in a CBC was quantified and shown to be negligibly small for the current gravitational wave detectors \cite{Adlk2}.\\
 In addition, one can use the analogous balance laws for angular momentum.  Work is in progress to calculate the posterior distributions for the spin of the remnant black hole at the end of a binary merger using the Phenom and EOB waveforms for events in the GWTC-1 catalog \cite{akk}.  As in \cite{adlkk},  the expectation is that a comparison between the two sets will serve as indicators of differences in the underlying physics.

Important background material is collected in the Appendix.  Although it is well known in the mathematical relativity literature, it is included here  to explain to non-experts why `supertranslations' must be included in the symmetry group of asymptotically Minkowskian spacetimes, since their presence leads to the supermomentum balance laws \cite{rg,aams} from which our constraints arise. This material is also used in the angular momentum considerations, discussed in \cite{Adlk2}, where the BMS symmetries and constraints discussed in this paper play a crucial role.

\section{Constraints on waveforms}
\label{s2}
This section is divided into  three parts. In the first, we introduce the basic framework that will be used to specify fields at $\scrip$. In the second, we introduce the notion of the Bondi--Metzner--Sachs (BMS) supermomentum, and the associated balance equations. In the third, we use these balance laws together with the standard asymptotic conditions in the CBC literature to arrive at an infinite family of constraints on waveforms.

\subsection{Underlying framework} 
\label{s2.1}

Our conventions are as follows. We work with $(-,+,+,+)$ signature and define curvature tensors via $2 \nabla_{[a} \nabla_{b]}K_{c} = R_{abc}{}^{d} K_{d}; \, R_{ac} = R_{abc}{}^{b}$ and $R= g^{ab} R_{ab}$.  We will assume that the physical spacetime $(M, g_{ab})$ is asymptotically Minkowskian in the sense spelled out, e.g., in \cite{aa-yau}. \footnote{This notion is weaker than Penrose's original definition of asymptotic simplicity which requires that every null geodesic in $M$ should have endpoints on $\scri^{\pm}$; our conditions refer only to properties of spacetime geometry near infinity.}

As explained in the Appendix, future null infinity, $\scrip$, is the natural home for all asymptotic fields. It is a null  3-manifold with topology $\mathbb{S}^{2}\times \mathbb{R}$, coordinatized by retarded time $u$ and angular coordinates $(\theta,\varphi)$. One can think of $\scrip$ either as the future boundary of the conformally completed spacetime $(\h{M} = M\cup \scrip,\, \hat{g}_{ab} = \Omega^{2} g_{ab})$ \`a la Penrose \cite{rp}, or as the limiting surface `$r = \infty$' obtained by moving away from sources along $u={\rm const}$ null surfaces (of constant retarded time) \`a la Bondi and Sachs \cite{bondi,sachs1,bondi-sachs}.  Since waveforms  are generally expressed using fields in the physical spacetime without making the conformal completion, in this paper we will do the same.

Since $(M, g_{ab})$ is asymptotically Minkowskian, following Bondi and Sachs let us introduce a foliation of the asymptotic region of $M$ by outgoing null hypersurfaces $u = {\rm const}$ and denote its geodesic null normal by $\l^{a}$. Introduce an affine parameter $r$ of $\l^{a}$ such that each null surface $u = {\rm const}$ is foliated by a family of (space-like) 2-spheres $r={\rm const}$. Denote the intrinsic $(+,+)$ metric of these 2-spheres by $q_{ab}$ and the other null-normal to each of these 2-spheres  by $n^{a}$, normalized so that $g_{ab} \l^{a} n^{b} = -1$. Finally, introduce a null complex vector field $m^{a}$ and its complex conjugate $\b{m}^{a}$ such that their real and imaginary parts are tangential to these 2-spheres, and they are normalized such that $g_{ab} m^{a} \b{m}^{b} =1$. Thus,  at each point in the asymptotic region we have a null tetrad  $\l^{a}, n^{a}, m^{a}, \b{m}^{a}$ for which the only non-zero contractions are $\l\cdot n =-1$ and  $m\cdot\b{m} = 1$. Finally, asymptotic conditions imply that this structure can be set up in such a way that, as $r\to \infty$, we acquire certain smooth  fields on $\scrip$, denoted here by a over-circle:

\begin{enumerate}[label=(\roman*)]
\item $\qo_{ab} = \displaystyle{\lim_{r\to\infty}} \,\,r^{-2}\, q_{ab}$\,  is an unit, round 2-sphere metric (and thus independent of  \indent  $u$);
\item $\no^{a} = \displaystyle{\lim_{r\to\infty}}\, n^{a}$ coincides with the null normal $\partial/\partial u$ to $\scrip$;
\item $\lo^{a} = \displaystyle{\lim_{r\to\infty}} \, r^{2} \l^{a}$ is the other null normal to a family of 2-sphere cross-sections $u={\rm const}$ of $\scrip$, and 
\item $\mo^{a} = \displaystyle{\lim_{r\to\infty}} \, r\,m^{a}$  and $\mbo^{a}=\displaystyle{\lim_{r\to\infty}} \, r\,\b{m}^{a} $ are tangential to these cross-sections.
\end{enumerate}
\vskip0.15cm

\begin{figure}[t]
\center
\includegraphics[width=.7\textwidth]{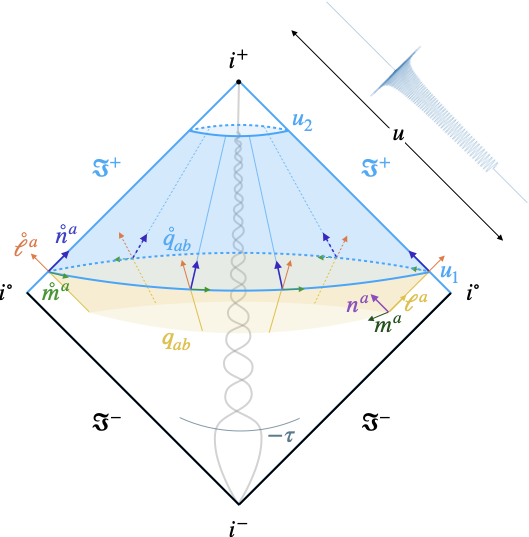} 
\caption{A depiction of future null infinity, $\scrip$ (in light blue) that constitutes the future boundary of space-time and has topology $\mathbb{S}^{2}\times \mathbb{R}$. The (yellow) outgoing null surface $u=u_{1}$ in the asymptotic region of space-time intersects $\scrip$ in a 2-sphere cross-section. The null tetrad $(\ell^a, n^a, m^a, \bar{m}^{a})$ and the metric $q_{ab}$ on $r={\rm const}$ 2-spheres in the interior have limits  $\lo^{a},\no^{a}, \mo^{a}, \mathring{\bar{m}}^{a}$ and $\qo_{ab}$ respectively on $\scrip$. The figure also shows an artist's impression of a typical waveform based on \cite{bohe2017improved}. }
\label{fig:scrip}
\end{figure}

In Penrose's conformal completion, one can always make the null normal $\no^{a}$ to $\scrip$ divergence-free by an appropriate choice of $\Omega$, and for each such choice, we acquire a pair $(\h{q}_{ab}, \h{n}^{a})$ of fields defined intrinsically on $\scrip$, with $\h{q}_{ab}$ the (degenerate) metric of signature $(+,+)$ and $\h{n}^{a}$ the null normal to $\scrip$, both induced  by the conformal metric $\h{g}_{ab}$. However, we can further restrict our conformal factor such that $\h{q}_{ab}$ is a \emph{unit 2-sphere metric}. The restricted pair $(\h{q}_{ab}, \h{n}^{a})$ is said to provide a `Bondi (conformal) frame'. In terms of structures introduced in the last para in physical spacetime,  we have $\qo_{ab} = \h{q}_{ab}$ and $\no^{a} = \h{n}^{a}$ at $\scrip$. The structure introduced is depicted in Fig.~\ref{fig:scrip}.

We will therefore refer to the pair  $(\qo_{ab}, \no^{a})$ as a \emph{Bondi-frame} at $\scrip$.  A change in the initial set up of the last paragraph can lead to a change in the Bondi frame.  Physically, each Bondi-frame selects an \emph{asymptotic Lorentz frame} because the vector field $\no^{a}$ provided by the Bondi-frame is an asymptotic time translation (i.e., limit to $\scrip$ of an asymptotic time translation vector field in spacetime). Recall that  we have a 3-parameter family of time-translation Killing fields ---or Lorentz frames--- in Minkowski space. Therefore, one would expect that there would be 3-parameter family of Bondi-frames. This is indeed the case: Bondi-frames are related to one another by (asymptotic) boosts. If we perform a boost by 3-velocity $\vec{v}$, the initial Bondi-frame $(\qo_{ab}, \no^{a})$ transforms to 
$({\mathring{\tilde{q}}}_{ab}, \, {\mathring{\tilde{n}}^{a}})$ given by 
\be\label{boost}  {\mathring{\tilde{q}}}_{ab} = \omega^{2} \qo_{ab}, \,\,\,\,  {\mathring{\t{n}}^{a}}= \omega^{-1}\no^{a}\quad  {\rm with} \quad \omega=  \f{1}{\gamma \left(1 - \f{\vec{v}}{c}\cdot \h{x}\right)}\,, \ee
where $\gamma = (1- (v/c)^{2})^{-\f{1}{2}}$ is the standard Lorentz factor and $\h{x}$ the unit radial vector with components $(\sin\theta \cos\varphi,\, \sin\theta\sin\varphi,\, \cos\theta)$. 
For simplicity we will use the terms `Bondi-frame' and `asymptotic Lorentz-frame' interchangeably. 

Recall from special relativity that the Lorentz frame in which the 4-momentum $P_{a}$ of the system is purely time-like ---i.e., in which the 3-momentum $\vec{P}$ vanishes is referred to as the \emph{rest frame} of the system. In most of this paper we will work with the \emph{past rest frame} at $\scrip$, i.e., the frame in which  the Bondi 3-momentum, defined in Sect.~\ref{s2.2} below,  vanishes in the asymptotic past, as $u\to -\infty$. Finally, at one intermediate step we will also have to use the Bondi-frame in which the Bondi 3-momentum vanishes in the distant future ---i.e., $u\to \infty$. In a CBC  this is the asymptotic Lorentz frame in which the final black hole is at rest, and represents the \emph{future rest frame}. \vskip0.15cm 
 
 In terms of the null-tetrads we have: $g_{ab} = -2 \l_{(a} n_{b)} + 2 m_{(a} \b{m}_{b)} = -2 \l_{(a} n_{b)} + q_{ab}$ on the physical spacetime $M$, and $\qo_{ab} = 2 \mo_{(a} \mbo_{b)}$ on $\scrip$.  Of special interest is the shear\, $\sigma = - m^{a}m^{b} \, \nabla_{a} \l_{b}$ \,of\, $\l^{a}$ because its asymptotic value $\sigmaz$ is directly related to the gravitational waveforms. In Minkowski spacetime  we can choose the $u={\rm const}$  surfaces to be the light cones emanating from points of a time-like geodesic. Then $\l^{a}$ is shear-free. In asymptotically Minkowskian spacetimes, shear $\sigma$ of  $\l^{a}$ falls-off as $1/r^{2}$ and 
\be \sigmaz (u,\theta,\varphi) :=  - \lim_{r\to\infty}  r^{2}  \big(m^{a} m^{b} \nabla_{a} \l_{b} \big)\,\equiv\, \frac{1}{2}(\hpz + i \hcz)(u,\theta,\varphi)  \label{shear}  \ee
is well-defined on $\scrip$, where, in the last step we have expressed $\sigmaz$ in terms of the commonly used strains
$$\hpz \equiv \lim_{r\to\infty}\,\, r h_{+} \qquad {\rm and} \qquad \hcz \equiv \lim_{r\to\infty}\,\, r h_{\times}.$$
(Because of the `$m^{a} m^{b}$' factor,  $\sigmaz$ has spin weight +2, while the more commonly used combination  $\hpz - i \hcz = 2\sigmabz$ has spin-weight -2.) The shear tensor $\sigmaz_{ab}$ at $\scrip$, given by
\be \sigmaz_{ab} (u,\theta,\varphi) =  - \big(\sigmaz\, \mbo_{a} \mbo_{b} \,+\, \sigmabz\, \mo_{a} \mo_{b}\,\big) (u,\theta,\varphi), \ee
is a symmetric, traceless tensor field, transversal to the null normal to $\scrip$  ---i.e. satisfies $\sigmaz_{ab} \no^{a} =0$--- and has spin weight zero. It captures the two Transverse-Traceless or radiative modes of the gravitational field in full, non-linear GR. Its time derivative is the Bondi news tensor: 
\footnote{It has a clear-cut geometric meaning in the conformally completed spacetime: $N_{ab}$ is the conformally invariant part of the curvature of the intrinsic connection on $\scrip$ \cite{aa-rad}.}
\be \label{news1} N_{ab} := 2 \Lie_{\no} \sigmaz_{ab}  \equiv   2\dot \sigmaz_{ab}\, .  \ee
One often introduces a `news function' $N$ via
\be \label{news2}  N_{ab} =2 \big( N \mo_{a} \mo_{b} + \b{N} \mbo_{a} \mbo_{b}\,\big)\, , \ee
so that $N =  -\partial_{u} {\b\sigma}^{\circ}  \equiv - \dot{\b{\sigma}}^{\circ}$ has spin weight -2.  

Finally let us introduce the Newman--Penrose components of the Weyl tensor that also feature prominently in the discussion of gravitational waves produced by CBCs. Since the stress energy tensor (if any) of the system is of compact spatial support, curvature in the neighborhood of infinity is fully captured by the Weyl tensor. Because $(M, g_{ab})$ is asymptotically Minkowskian, components of the  Weyl tensor in the null tetrad have a specific fall-off, known as `peeling properties' \cite{sachs1,np}. The leading order components of the Weyl tensor in the null tetrad ---which carry a superscript $\circ$--- can be regarded as fields on $\scrip$:
\ba \label{weyl} 
\4z (u,\theta,\varphi)&= \displaystyle{\lim_{r\to\infty}\,  r C_{abcd} n^{a} \b{m}^{b} n^{c} \b{m}^{d}};\qquad 
\3z (u,\theta,\varphi)&= \lim_{r\to\infty}\,  r^{2} C_{abcd} n^{a} \b{m}^{b} n^{c}  {\l}^{d}; \nonumber\\
2\,\Re\,\2z (u,\theta,\varphi) &= \displaystyle{\lim_{r\to \infty}\,  r^{3} C_{abcd} n^{a} {\l}^{b} n^{c}  {\l}^{d}};\quad
 2\, \Im\,\2z (u,\theta,\varphi)&=   -i\,\lim_{r\to \infty}\,  r^{3} C_{abcd} n^{a} {\l}^{b} m^{c}  {\b{m}}^{d}; \nonumber\\
 \1z (u,\theta,\varphi)&= \displaystyle{\lim_{r\to \infty}\,  r^{4} C_{abcd} \l^{a} {m}^{b} \l^{c}  {n}^{d}}; \qquad
 \0z (u,\theta,\varphi)&= \lim_{r\to \infty}\,  r ^{5}C_{abcd} \l^{a} {m}^{b} \l^{c} {m}^{d}\,. \ea
The spin-weight  $-2$  component $\4z$  is referred to as the \emph{radiation field} because: (i) it is the leading order coefficient of curvature component with asymptotic fall-off $1/r$; and, (ii) it is directly related to the waveforms, namely \vskip0.3cm
\centerline{$\4z  =  - \ddot{\bar\sigma}^{\circ} \equiv  -\partial_{u}^{2} \sigmabz\, .$}
\vskip0.3cm
\noindent By contrast, the real part of the (spin-weight zero) component $\2z$ encodes the `Coulombic' information contained in the Bondi mass. The angular momentum information resides in the spin-weight $1$ component $\1z$ \cite{td,ak-thesis}. This may seem surprising at first because one is accustomed to the statement that in Kerr spacetime the only non-zero component of the Weyl tensor is $\Psi_{2}$. However, that statement refers to components in principal null directions which do not agree with the pair of null vector fields $(\no^{a}, \lo^{a})$ at $\scrip$.  In terms of the null tetrad at $\scrip$ in a Kerr spacetime with mass $M$ and angular momentum $J$,\,\, $\2z$ is a real constant with value $-GM$, while  $\1z = (3i/2) \sin\theta\, GJ$. Finally, in this paper we will not need the component $\0z$; it was introduced just for completeness.\\
 
 \emph{Remark}: The fall-off in $1/r$ of curvature components  in the physical spacetime is often referred to as ``peeling'' and translates to degree of differentiability  ---or smoothness--- of the conformally rescaled metric $\h{g}_{ab}$ in the Penrose completion. There has been considerable debate in the literature on whether these assumptions are physically reasonable (for a summary, see e.g. \cite{hf2}). Indeed, using the PN perspective it was argued in the early days that they may not be appropriate for CBC \cite{lbtd}. However, as noted in the Appendix, the current consensus in the PN literature is that the asymptotic form of the PN metric is completely consistent with the Bondi--Sachs--Penrose framework \cite{blanchet1}. Similarly,  the key  notions underlying this framework ---such as  $\4z$, $N$, and $\sigmaz$--- and relations between them are heavily used in NR.  
 
\subsection{The supermomentum balance law}
\label{s2.2}

Much of the previous work  on supermomentum is carried out in the conformally completed spacetime, without introducing auxiliary structures such as the tetrad vectors $\ell^{a}, m^{a}, \bar{m}^{a}$ that are not canonically defined at $\scrip$ \cite{rg,aams,aa-bib}. Similarly, one does not require that the intrinsic metric $\qo_{ab}$ on $\scrip$ is the round, unit 2-sphere metric. This generality made it manifest that the results have all the required invariance and covariance properties. However, since the waveforms community does not use conformal completion, we will now present the material using only those notions that are introduced in Sect.~\ref{s2.1} (and in the Appendix). On the other hand,  to check transformation properties, e.g., from one Bondi frame to another, it is easiest to use the original invariant framework.

As explained in the Appendix, one of the major surprises to emerge from the Bondi--Sachs work was that, although spacetimes under consideration are asymptotically Minkowskian, the asymptotic symmetry group is not the Poincar\'e group $\P$, but an infinite dimensional enlargement thereof, the BMS group $\B$. While $\B$ does admit a canonical 4-dimensional normal subgroup $\T$ of translations \cite{sachs2}, it also contains an infinite dimensional Abelian subgroup $\S$ of  \emph{supertranslations}, which can be thought of as `angle dependent translations'. There is a direct relation between this enlargement and the presence of gravitational waves: In absence of gravitational radiation the group naturally reduces back to the Poincar\'e group \cite{np2,aa-rad,aa-bib}. Intuitively, the enlargement occurs because the ripples of curvature propagate out all the way to infinity, introducing an ambiguity in the choice of the `background Minkowski metric' that the physical metric $g_{ab}$ is approaching as $1/r$. In any Bondi-frame at $\scrip$, we have the following properties:
\begin{enumerate}[label=(\roman*)]
\item translations are represented by vector fields $\xi^{a}_{(\alpha)} := \alpha(\theta,\phi) \no^{a}$ at $\scrip$, where $\alpha$ is a (real-valued) linear combination of the first 4 spherical harmonics: 
\be \label{alpha} \alpha(\theta,\varphi) = \alpha_{o} Y_{0,0}+\sum_{m=-1}^{m=1}\,  \alpha_{m} Y_{1,m} (\theta,\varphi)\ee 
for some constants $\alpha_{0}, \alpha_{m}$;
\item the vector field $\xi^{a}_{(0)} = \no^{a}$ is \emph{the} time-translation in the chosen Bondi frame and the vector fields $\xi^{a}_{(\vec\alpha)} =  \sum_{m=-1}^{m=1}\, \alpha_{m} Y_{1,m} (\theta,\varphi)\, \no^{a}$ are spatial translations in that frame;
\item the more  general vector-fields $\xi^{a} _{(f)}= f(\theta,\varphi)\, \no^{a}$ where $f$ is any smooth function on a 2-sphere are called supertranslations; so translations are just special cases of supertranslations. If we fix a Bondi-frame ---as we will in most of the paper, taking it to be the rest-frame in the asymptotic past--- we can speak of `pure' supertranslations: These correspond to
\be f(\theta,\varphi)  = \sum_{\l =2}^{\infty}\, \sum _{m=-\l}^{\l}\, f_{\l,m} Y_{\l,m} (\theta,\varphi)\, .\ee 
\end{enumerate}
In what follows, we will use the symbol $\alpha(\theta,\varphi)$ to refer to BMS translation and $f(\theta,\varphi)$ to refer to generic BMS supertranslations.%
\footnote{\label{fn2}  If we change the Bondi-frame, the vector field $\no^{a}$ is rescaled as in Eq. (\ref{boost}). Since the supertranslation is given by $\xi^{a} = f \no^{a} = \t{f} \mathring{\t{n}}^{a}$, the labels $f$ and $\t{f}$ in the two frames are related by $\t{f} = \omega f$. In the Penrose conformal picture, $f$ is not a scalar but carries a conformal weight $1$. It turns out that the notion of a BMS translation is invariant with respect to this change of the Bondi-frame, but the notion of  a `pure supertranslation' is not.}

Recall that for fields in Minkowski space, energy-momentum arises as the Hamiltonian generating  canonical transformations corresponding to spacetime translations. In the same spirit, in GR one can construct a phase space $\Gamma_{\rm Rad}$ of radiative modes at $\scrip$ \cite{aaam2}, show that the action of BMS supertranslations $\xi^{a}_{(f)}$ on $\scrip$ induces canonical transformations on $\Gamma_{\rm Rad}$, and calculate the corresponding Hamiltonians \cite{aams}. They are given by integrals over $\scrip$ which have the interpretation of the \emph{total fluxes}  $\F_{(f)}$ across $\scrip$ of the components of supermomentum defined by $f(\theta,\varphi)$:
\ba \label{flux1} \F_{(f)}  &=&  \f{1}{32\pi G} \int_{\scrip} \rmd u\, \d2vo\, f(\theta,\varphi) \, \big[ N_{ab}N^{ab} + 2 \Do_{a}\Do_{b} N^{ab}\big] (u,\theta,\varphi) \nonumber\\
&=& \f{1}{4\pi G} \int_{\scrip} \rmd u\, \d2vo\, f(\theta,\varphi) \, \big[ |\dot\sigmaz|^{2} - \Re(\eth^{2} \dot{\bar\sigma}^{\circ}) \big] (u,\theta,\varphi)\, . \ea
Here, in the first step we have raised the indices of $N_{ab}$ using the unit 2-sphere metric $\qo^{ab}$ on the $u={\rm const}$ 2-spheres, and $\Do_{a}$ is the derivative operator compatible with $\qo_{ab}$. In the second step we have used the Newman--Penrose angular derivative operator $\eth$ whose action on a spin-weight $s$ scalar $A$, given by
\be \label{eth} \eth  A = \frac1{\sqrt{2}}{(\sin\theta)^{s}} \left(\f{\partial}{\partial\theta}\, +\, \f{i}{\sin\theta}\, \f{\partial}{\partial \varphi}\right) (\sin\theta)^{-s} A  \, ,\ee
yields a scalar with spin-weight $s+1$. (In our case, $\dot{\bar\sigma}^{\circ}$ has spin-weight $-2$, whence $\eth^{2} \dot{\bar\sigma}^{\circ} $ has spin-weight zero.  Please note that in the literature, $\eth$ often carries the opposite sign relative to Eq. (\ref{eth}).) The second equality in (\ref{flux1}) provides us with supermomentum fluxes  directly in terms of the waveform $2\sigmaz (u,\theta,\varphi)= (\hpz + i \hcz)(u,\theta,\varphi)$.

One can show that the $\F_{(f)}$ is in fact an integral of an exact 3-form, whence the integral on $\scrip$ can be expressed as the difference between two 2-sphere integrals, performed at $\iz$ (i.e. on the `$u= -\infty$' 2-sphere) and $i^{+}$ (i.e. on the `$u= \infty$' 2-sphere) \cite{aams}:
\be \label{bal} \F_{(f)} = \lim_{u_{o} \to -\infty}\, P_{(f)}|_{u=u_{o}} - \lim_{u_{o} \to \infty}\, P_{(f)}|_{u=u_{o} }\ee
where 
\be \label{supmom} P_{(f)}\big{|}_{u=u_{o}}  :=  - \f{1}{4\pi G} \,\oint_{u=u_{o}}\!\! \!\!\d2vo\, f(\theta,\varphi)\, \Re \big[\2z + \sigmabz \dot\sigmaz\big](\theta,\varphi)\,  \ee
is the $f$-component of the supermomentum of the system at the retarded time $u=u_{o}$. 
Equation~(\ref{bal})  constitutes an infinite family of \emph{balance laws} ---one for each choice of a supertranslation, i.e., of  smooth $f(\theta,\varphi)$ on a 2-sphere.  As we will show in  Sect.~\ref{s2.3},  these laws provide us with an infinite family of constraints on waveforms. Finally, Refs. \cite{rg,aams} provide more general expression of supermomentum that hold on \emph{any} 2-sphere cross-section of $\scrip$ (not necessarily $u=\rm{const}$) and formula of supermomentum flux through \emph{any} patch $\Delta\scrip$ of $\scrip$.  Eqs (\ref{flux1}) and (\ref{supmom}) are just special cases of  those expressions adapted to the setup introduced in Sect.~\ref{s2.1}.

Since the balance law holds for any supertranslation $f(\theta, \varphi) \no^{a}$, it holds in particular for translations $\alpha(\theta,\varphi) \no^{a}$. In this case $P_{(\alpha)}$ is simply the Bondi--Sachs energy-momentum of the system at the retarded time $u=u_{o}$.  Since the functions $\alpha(\theta,\varphi)$ are linear combinations of the first four spherical harmonics, the flux formula simplifies.  For, $\qo^{ab}\Do_{b} \alpha$ is a conformal Killing field of the 2-sphere metric $\qo_{ab}$, whence $\Do_{a} \Do_{b} \alpha\, \propto\, \qo_{ab}$. Therefore, integrating the first equation in (\ref{flux1}) by parts and using the fact that $N_{ab}$ is trace-free, one obtains:
\be \label{flux2} \F_{(\alpha)}  =  \f{1}{4\pi G} \int_{\scrip} \rmd u\, \d2vo\, \alpha(\theta,\varphi) \, |\dot\sigmaz|^{2}(u,\theta,\varphi)\, .  \ee
In particular, for any \emph{time}-translation, $\alpha(\theta,\varphi)$ is positive, whence the flux of energy is necessarily positive. This is the celebrated Bondi--Sachs result.  To calculate the final black hole kick, one computes the 3-momentum flux $\F_{(\vec{\alpha})}$, where the function $\vec{\alpha}$ is a general linear combination of the three $Y_{1,m}$. \\ 

\emph{Remark:}  If one accepts  (\ref{supmom}) as the definition of supermomentum, then the balance law (\ref{bal}) follows directly from Einstein's equations and Bianchi identities. However, without the use of Hamiltonian methods, it is difficult to justify why this expression should be interpreted as supermomentum. Indeed,  in the literature between 1960s and early 1980s there was considerable confusion about supermomentum  because the flux expressions were not obtained from a physically well-motivated procedure. For example, expressions of supermomentum given in \cite{jw,bramson,crp,ms} lead to a non-zero flux  between general cross-sections of $\scrip$ \emph{in Minkowski space} \cite{rgjw,aajw,ts}! The expression (\ref{flux1}) by contrast vanishes anytime the news tensor vanishes, in particular in any stationary spacetime (relation between these flux expressions is discussed in \cite{aajw}). Even when Hamiltonian considerations were used to provide a physical basis for the derivation, the flux expression in the early literature was incorrect (see, e.g.  second article in \cite{bondi-sachs} and \cite{ms-thesis}) because of a subtle fact that the radiative phase space is an affine space, rather than a vector space (for details, see \cite{aams}). These examples serve to bring out the fact that considerable care is needed to arrive at viable balance laws.

\subsection{Implications for waveforms}
\label{s2.3}

We can now apply the balance laws to isolated systems undergoing CBC. Since the balance laws refer to exact GR, we will not need to make any approximations. However, we need to incorporate the fact that we are now restricting ourselves to CBC. We will do so through two physically motivated assumptions in Sect.~\ref{s2.3.1}. The resulting constraints and their possible applications on the global waveforms are discussed in Sect.~\ref{s2.3.2}, while Sect.~\ref{s2.3.3} discusses the constraints on restrictions of waveforms to finite $u$ intervals.

\subsubsection{Assumptions}
\label{s2.3.1}

\begin{enumerate}[label=\emph{(\arabic*)}]
\item \emph{We assume that the Bondi news tensor $N_{ab}$ on $\scrip$ goes to zero as $u\to \pm \infty$ as $1/u^{1+\epsilon}$ for some $\epsilon >0$}.\!
\footnote{Throughout, we assume that if a field $F(u,\theta,\phi) = O(1/|u|^{\alpha})$\,\, ---i.e., if  $|u|^{\alpha}F(u,\theta,\varphi)$ admits smooth limits $F_{\pm}(\theta,\varphi)$ as $u\to \pm\infty$--- \,\,then its $m$th\, $u$-derivative, $\partial_{u}^{m} F(u,\theta,\varphi)$  is $O(1/|u|^{m+\alpha})$.}
\label{as:news}
\end{enumerate}
This condition is necessary and sufficient to ensure that the flux of  the BMS angular momentum across $\scrip$ is finite \cite{aams,Adlk2}.  For vacuum solutions, the Christodoulou-Klainnerman results, for example, ensure that this condition holds with $\epsilon =1/2$. Waveforms constructed so far satisfy this assumption because the Bondi news goes to zero rapidly at early and late times; within the approximations made and accuracy achieved, $N_{ab}$ is in fact indistinguishable from zero outside some finite $u$-interval. In terms of the waveform $\sigmaz = \hpz + i \hcz$, our assumption reads:
\be \label{assum1}  \sigmaz (u,\theta,\varphi) = \sigma_{\pm} (\theta,\varphi) + |u|^{-\epsilon} \, \sigma^{(1)}_{\pm} (\theta,\varphi) + O (|u|^{-\epsilon-1} ) \quad {\rm as}\,\, u\to \pm\infty \,,\ee
\vskip0.15cm
\noindent so that $\sigma_{\pm} (\theta,\varphi)$ are the limits of $\sigmaz (u,\theta,\varphi)$, and $\pm\epsilon \sigma^{(1)}_{\pm} (\theta,\varphi)$ are the limits of $|u|^{1+\epsilon}\,\bar{N}(u,\theta,\varphi)$ as $u \to \pm \infty$.

 Next, it is expected that the system would settle down to a stationary state after  coalescence. Indeed, in the case of binary black holes, numerical simulations show that at late times the spacetime metric approaches the Kerr solution. In addition, the PN calculations assume that the system is stationary in the past, i.e., for times $t < -\tau$ \cite{blanchet1}. We will need a much weaker assumption to capture the physical behavior of  the compact binary in the asymptotic future and past.  
 
Let us first note a consequence of assumption \ref{as:news}.  Since $N_{ab} = O(1/|u|^{1+\epsilon})$ as $u\to \pm\infty$, Einstein's equations together with Bianchi identities imply that $\partial_{u} \4z, \,\,\partial_{u}\3z$ and $\partial_{u}\2z$ go to zero as $u^{-(3+\epsilon)}, \,\,u^{-(2+\epsilon)}$ and $u^{-(1+\epsilon)}$ respectively. In this precise sense these three complex components of the asymptotic curvature become  `time-independent'  or `stationary' \emph{in the limit} $u\to \pm \infty$.  Because $N_{ab}$ is invariant under the change of Bondi-frame, this fall-off behavior holds in \emph{any} Bondi-frame. Furthermore,  $\4z$ and $\3z$ themselves vanish in these limits, and for $\2z$ we have 
\be \label{psi2}  \2z (u,\theta,\varphi) = \psi_{\pm}(\theta,\varphi) + |u|^{-\epsilon} \psi_{\pm}^{(1)} (\theta,\varphi) + O (|u|^{-\epsilon-1} ) \quad {\rm as}\,\, u\to \pm\infty \, ,\ee
again in any Bondi-frame, although the limiting values $\psi_{\pm}(\theta,\varphi)$ and $\psi_{\pm}^{(1)} (\theta,\varphi)$ depend on the choice of the frame.  Note, however, that  assumption \ref{as:news} does \emph{not} imply that $\1z$ (or $\0z$) become stationary as $u\to \pm \infty$. Furthermore,  the limiting values of $\partial_{u} \1z$ depend on the choice of the Bondi-frame and in general the rest-frame of the final black hole is different from that of the system in the distant past because of black hole kicks \cite{kicks1,kicks2}.  We will now make our second assumption: 
\begin{enumerate}[label=\emph{(\arabic*)},resume]
\item \emph{$\partial_{u} \1z \to 0$ in the past  Bondi-frame as $u\to -\infty$, and in the future Bondi-frame as  $u\to \infty$.} \label{as:psi1}
\end{enumerate}
Recall that the past Bondi-frame is the one in which the Bondi 3-momentum vanishes in the distant past (i.e., as $u\to -\infty$), and the future Bondi-frame is the one in which the Bondi 3-momentum vanishes in the distant future (i.e., as $u\to\infty$). Thus,  we are \emph{not} requiring that the system should become stationary in the past and the future in the same rest-frame.  As we discuss below, that requirement would have been too restrictive. Ours is quite weak: in the distant past it is \emph{much weaker} than what is assumed in PN waveforms, and it is found to hold in the NR simulations in the distant future, since the solution quickly relaxes to a stationary spacetime \cite{Boyle:2019kee}.

\subsubsection{Constraints on global waveforms}
\label{s2.3.2}

The two assumptions lead to a key simplification in the surface terms representing asymptotic values of supermomentum (\ref{supmom}) used in the balance laws (\ref{bal}).  First, in any Bondi-frame, Einstein's equations and Bianchi identities imply that the following relations hold on all of $\scrip$ (see, e.g., \cite{np}):
\ba \label{1dot}  \partial_{u} \1z  &=& m^{a} \Do_{a} \2z + 2 \sigmaz\, \3z \,\, \equiv\,\, \eth\2z - 2 \sigmaz\, \eth \dot{\bar{\sigma}}^{\circ} , \quad {\rm and,} 
\\
\label{impsi2} \Im\, \2z &=&  \Im\, \big(\b{\eth}^{2}\, \sigmaz +  \sigmaz\, N\,\big) \, .
\ea
They have two important consequences. Let us first consider the \emph{past} Bondi-frame.
Then,
\begin{enumerate}[label=(\roman*)] 
\item in the limit $u \to -\infty$ the left side of (\ref{1dot}) vanishes by assumption \ref{as:psi1}, while the second term on the right hand side vanishes by assumption \ref{as:news}. Therefore $m^{a} \Do_{a} \2z$ vanishes ---i.e., \emph{$\2z$ becomes spherically symmetric}--- in the limit $u\to -\infty$. \\ 
\item Next, thanks to assumption \ref{as:news}, Eq. (\ref{impsi2})  implies\,\, $ \lim_{u\to -\infty} \,\Im\, \2z  = \b{\eth}^{2}\, \Im\, \sigma_{-}$. Integrating both sides over a 2-sphere, and using spherical symmetry of $\Im\,\2z$ we conclude that $\Im\, \2z = 0$ in the limit.
\end{enumerate}

 To evaluate the limit of $\Re\,\2z$, we will use the expression (\ref{supmom}) to calculate the past limit of the Bondi 4-momentum.  This limit  is purely time-like in the rest-frame at $\iz$,  and the limiting Bondi energy is precisely the initial (or the total) mass $M_{\iz}$. Therefore, setting $f=1$ in  Eq. (\ref{supmom}), and taking the limit $u_{o} \to -\infty$ we obtain:
 \begin{enumerate}[label=(\roman*),resume]
\item  $\lim_{u\to -\infty} \,\Re\, \2z = - GM_{\iz}$. 
\end{enumerate}

Thus, we conclude:
\be \label{psi2o}  \displaystyle{\lim_{u\to -\infty}} \, \Re\, \2z (u,\theta,\varphi)  = -GM_{\iz},  \quad {\rm and} \quad \displaystyle{\lim_{u\to -\infty}} \, \Im\, \2z (u,\theta,\varphi)  =0\, .   \ee

Now, the asymptotic rest frame of the system at $u=\infty$ is generically different from that in the past. To work out the implication of condition \ref{as:psi1} on the behavior of $\2z$ in the distant future, let us for a moment switch to the future Bondi-frame ---in which the \emph{final} black hole is at rest--- and denote fields in that frame with a prime. Then the reasoning we used also implies that
\be  \label{past2z} \lim_{u\to\infty}\,{\2z}^{\prime} = - GM_{i^{+}}.\ee 
Since the balance law is formulated in a general but \emph{fixed} Bondi-frame, and we have chosen to work in the past Bondi-frame, we need to transform ${\2z}^{\prime}$ to that frame. The two frames are related to each other by a boost defined by the velocity $\vec{v}$ of the final black hole (in the past rest-frame),  commonly referred to as the \emph{kick} velocity.  
To calculate the components $v_{i}  = (v_{x},\, v_{y},\, v_{z})$  of $\vec{v}$ in the past rest frame, we first note that since initially the Bondi 3-momentum is zero, the 3-momentum $P_{(\alpha_{i})}$ of the final black hole in the past rest-frame is just the negative of  the Bondi 3-momentum flux $\F_{(\alpha_{i})}$ carried by gravitational waves, and hence completely determined by the waveform. From Eq. (\ref{flux2}) we obtain
\be 
\gamma\,\,M_{i^{+}}\,\,v_{i} \equiv P_{(\alpha_{i})} \,\,\,=\,   - \f{1}{4\pi G} \int \rmd u\, \d2vo\, \alpha_{i}(\theta,\varphi) \,\, |\dot\sigmaz (u,\theta,\varphi)|^{2}\,  , 
\ee
{\sloppy where, again $\gamma = (1 -v^{2}/c^{2})^{-\f{1}{2}}$ is the standard Lorentz factor and $\alpha_{i}(\theta,\varphi) = (\sin\theta\cos\varphi,\,\sin\theta\sin\varphi,\, \cos\theta) \equiv \h{x}$. In terms of this velocity $\vec{v}$, the transformation property of $\2z$ under the change of Bondi-frames implies that  the future limit of $\2z$ in the past Bondi-frame is given by} 
\footnote{This transformation property follows from the conformal rescaling of $\no^{a}$ given in Eq. (\ref{boost}), the normalization condition $g_{ab}\ell^a n^b=-1$ in physical spacetime, the definition of $\2z (u,\theta,\phi)$ given in Eq. (\ref{weyl}) and the fact that  the radial coordinate changes via $r'=\gamma\,(1-\tfrac{\vec{v}}{c}\cdot\h{x})\, r +O(1)$ under a boost.}
\be \label{trans} 
 {\2z} \big{|}_{u=\infty} \,=\, \f{{\2z}^{\prime}}{\gamma^{3}\,\big(1 - \f{\vec{v}}{c}\cdot \h{x} \big)^{3}}  \, =\,   -\, \f{GM_{i^{+}}}{\gamma^{3}\,\big(1 - \f{\vec{v}}{c}\cdot \h{x} \big)^{3}} \, .  \ee 
Thus, using assumptions \ref{as:news} and \ref{as:psi1} we have now expressed  $\2z$ in both asymptotic limits $u\to \pm\infty$ using the the past Bondi-frame through Eqs. (\ref{psi2o}) and (\ref{trans}). \\

We can now use the balance law (\ref{bal}) for a general supertranslation $f(\theta,\varphi)$, in conjunction with with the forms (\ref{past2z}) and (\ref{trans})  of $\2z$ in the asymptotic past and future, to conclude that the waveform $\sigma^\circ = (\hpz + i \hcz)/2$ must satisfy
\be \label{smearedkey1} \oint \d2vo\, f(\theta,\varphi) \Big(GM_{\iz}  -  \f{GM_{i^{+}}}{\gamma^{3} \big(1- \f{\vec{v}}{c} \cdot \h{x}  \big)^{3} }\,\Big) \,\,=\,\,  \oint \d2vo\,  f(\theta,\varphi) \int_{-\infty}^{\infty}\!\!\rmd u\,  \big[ |\dot\sigmaz|^{2} - \Re(\eth^{2} \dot{\bar\sigma}^{\circ}) \big] (u,\theta,\varphi)\, . \ee
Since $f(\theta,\varphi)$ is an arbitrary smooth function, we can peel-off the 2-sphere integral to obtain
\be \label{key1}  GM_{\iz}  -  \f{GM_{i^{+}}}{\gamma^{3} \big(1- \f{\vec{v}}{c} \cdot \h{x} \big)^{3} }\,\,=\,\,  \int_{-\infty}^{\infty} \rmd u\,  \big[ |\dot\sigmaz|^{2} - \Re(\eth^{2} \dot{\bar\sigma}^{\circ}) \big] (u,\theta,\varphi)\,\, =: \,\F(\theta,\varphi)\,.  \ee
$\F(\theta,\varphi)$ can be thought of as the `total flux of supermomentum in the $(\theta,\varphi)$ direction'. Since, Eq.(\ref{key1}) holds for each $(\theta,\varphi)$ \emph{we have an infinite set of conditions.} 

Now, every waveform used for data analysis provides us with:
\begin{enumerate}[label=(\roman*)]
\item the waveform $\sigmaz(u,\theta,\varphi) = \f{1}{2} (\hpz+ i \hcz)$ that completely determines the right side of (\ref{key1}), as well as the kick velocity $v$ through Eq. (\ref{flux2});
\item the total initial mass of the system $M_{\iz}$;  
\item the mass $M_{i^{+}}$ of the final black hole.
\end{enumerate}
If the global procedure used to create the waveform is to be consistent with exact GR, then the outcome must satisfy the infinite set of constraints (\ref{key1}).  This is our principal observation.  \vskip0.15cm 

To get an intuitive feel for these constraints, let us first consider the special case where there is no kick, i.e.,  where  the velocity $v$ of the final black hole is zero. Then the left side of (\ref{key1}) is a constant, while the right side $\F(\theta,\varphi)$, constructed from the waveform, is a function of $(\theta,\varphi)$.  The infinite set of constraints now imply that $\F(\theta,\varphi)$ must be a constant. However, zero kick represents a very special CBC. Let us now consider the general case. Numerical simulations show that typical kick velocities are a few hundred km/s. For definiteness, let us take it be  $\sim 300$ km/s, so $v/c\, \sim\, 10^{-2}$.  One can then Taylor expand the second term in the left side of (\ref{key1})  in $v/c$, 
\be \label{key1approx}  GM_{\iz}\,  - \, \f{GM_{i^{+}}}{\gamma^{3} \big(1- \f{\vec{v}}{c} \cdot \h{x}  \big)^{3}}  \,=\, GM_{\iz}\, -\,GM_{i^{+}}\,\Big(1 + \big(3\,\h{v}\cdot \h{x}\big)\, \f{v}{c} - \big(\f{3}{2} - 6 (\hat{v}\cdot \h{x})^{2} \big) \, \f{v^{2}}{c^{2}} + \ldots \Big) \ee
and keep terms to the desired accuracy (here $\hat{v}$ is the unit vector in the $\vec{v}$ direction). Within this scheme, to test if a given waveform is accurate to  $\sim0.3\%$ we only need to retain the first order term in $v/c$, namely,  $(-3\,GM_{i^{+}}\,\, \h{v}\cdot\h{x})\, v/c$. Then the right side has only linear combinations of $\ell =0,1$ spherical harmonics, whence, in the expansion of $\F(\theta,\varphi)$  in terms of spherical harmonics, \emph{all coefficients of the flux must vanish for $\ell \ge 2$}. This means that the waveform must be such that the total flux of the infinitely many components of  \emph{pure} supermomentum  across $\scrip$ must vanish to $\sim 0.3\%$ accuracy. If we are interested in $0.01\%$ accuracy, we would keep terms up to $O(v^{2}/c^{2})$ and  should then find that the coefficients of the total flux must vanish for $\l>2$, and so on. \emph{To summarize, (\ref{key1approx}) provides an infinite number of conditions that any proposed waveform must satisfy if it  is to approximate the outcome of the exact GR calculation to a given desired accuracy. }  The resulting error-bars are relative to exact GR rather than NR simulations (which come with their own error bars).

\subsubsection{Constraints on waveforms over finite time intervals}
\label{s2.3.3}

So far we have focused on the `global balance law' (\ref{bal}) that refers to all of $\scrip$. However, we also have a `quasi-local' balance law for any portion $\Delta\scrip$ that is bounded by two cross-sections $C_{1}$ and $C_{2}$ \cite{rg,aams}. For waveform considerations it suffices to take $C_{1}$ and $C_{2}$ to be $u=u_{1}(\theta,\varphi)$ and $u=u_{2}(\theta,\varphi)$ cross-sections in any given Bondi-frame. Then we have:
\be \label{bal2}   P_{(f)}|_{u=u_{1}} -  P_{(f)}|_{u=u_{2} } = \F_{(f)}\big{|}_{u_{1}}^{u_{2}} \ee
where the supermomenta $P_{(f)}|_{u_{i}}$  (with $i =1,2$) on the left hand side are again given by (\ref{supmom}),
\be  P_{(f)}\big{|}_{u=u_{i}}  :=  - \f{1}{4\pi G} \,\oint_{u=u_{i}}\!\! \!\!\d2vo\, f(\theta,\varphi)\, \Re \big[\2z + \sigmabz \dot\sigmaz\big](\theta,\varphi)\,\ee
and the flux on the right side is given by
\be \label{flux3} 
\F_{(f)}\big{|}_{u_{1}}^{u_{2}} =  \f{1}{4\pi G} \oint\d2vo\, f(\theta,\varphi) \,
\int_{u_{1}}^{u_{2}}\!\! \rmd u\, 
 \big[ |\dot\sigmaz|^{2} - \Re(\eth^{2} \dot{\bar\sigma}^{\circ}) \big] (u,\theta,\varphi)\, . \ee
One can again `peel off' the 2-sphere integral using the fact that $f(\theta,\varphi)$ is arbitrary and arrive at the  local analog of (\ref{key1}):
\be \label{key2}  \Re \big[\2z + \sigmabz \dot\sigmaz\big](u_{1}, \theta,\varphi) - \Re \big[\2z + \sigmabz \dot\sigmaz\big](u_{2 },\theta,\varphi)  
=   - \int_{u_{1}}^{u_{2}}\!\! \rmd u\, 
 \big[ |\dot\sigmaz|^{2} - \Re(\eth^{2} \dot{\bar\sigma}^{\circ}) \big] (u,\theta,\varphi)  \ee

The quasi-local constraint (\ref{key2}) has richer content than the the global constraints \eqref{key1}: since $u_{1}$ and $u_{2}$ are arbitrary, it provides dynamical checks on the waveform \cite{tdl-aps}. However, in order to verify  whether it is satisfied, one needs to know $\2z$. The SXS collaboration is expected to release simulations that extract $\2z$ in the NR regime in the very near future \cite{lk,di}, making (\ref{key2}) a potentially powerful tool to estimate errors in the simulations vis a vis exact GR. In addition, as noted in Sect.~\ref{s1},   the `finite time balance law' (\ref{key2}) has already been used to improve certain modes of waveforms in the SXS catalog \cite{mikbdLkmpst}. 

We will conclude with a few remarks on Sect.~\ref{s2} as a whole.
\begin{enumerate}
\item Had spacetime been exactly stationary for $t < - \tau$,  as is generally assumed in the PN literature \cite{lbtd,blanchet1,bi,ep}, then we could have used the multipolar expansion of the metric that is shown to hold rigorously for stationary spacetimes outside some spatially compact region \cite{rbws}. Using this expansion, one can calculate the asymptotic Weyl tensor and show that  $\2z$ is real and spherically symmetric on $\scrip$  in the past Bondi-frame on \emph{an infinite} interval  $u < -u_{o}$  for some $u_{o}$.  (In this case, the past Bondi-frame is the one in which $\no^{a}$ is the limit to $\scrip$ of the stationary Killing field.) Since our notion of asymptotic stationary in the past is extremely weak in comparison, we can only conclude that $\2z$ becomes spherically symmetric in the limit $u\to -\infty$.

\item All analytical approximations used to construct waveforms ---such as EOB--- agree with the PN one in the distant past.  Since our weak assumption \ref{as:psi1} is inspired by the PN stationarity condition, it holds in these approaches as well.  

\item Since the Lorentz transformation of energy only involves the first power of $\gamma (\vec{v}\cdot \h{x}/c)$ in the denominator, the appearance of the cube of this factor in (\ref{key1}) may seem surprising. The additional factor of  $(\gamma (\vec{v}\cdot \h{x}/c))^{-2}$ comes from the transformation property of the 2-sphere area-element: We have `peeled-off' a 2-sphere integral in the passage from (\ref{bal}) to (\ref{key1}).

\item If there is no kick, i.e., if $v=0$ identically, then the past and future Bondi frames coincide.   This is of course a \emph{very special case} realized, e.g., in an equal mass head-on collision. In this case, Eq. (\ref{trans}) simplifies to $\2z|_{u=\infty} = - GM_{i^{+}}$. Thus $\2z$ becomes spherically symmetric \emph{in the same Bondi-frame} in both asymptotic limits, $u\to \pm \infty$. Therefore, in the terminology used in the literature, the \emph{linear (or, ordinary) memory vanishes}.  Recently, this implication of asymptotic stationarity of the system in the \emph{same Bondi frame,} as $u\to \pm\infty$, was arrived at  using detailed calculations using the physical spacetime metric \cite{gsrw}.  Analysis at $\scrip$ summarized above brings out the minimal setup needed to obtain this result:
\begin{enumerate}[label=(\roman*)] 
\item One only needs that the past and future rest frames are the same, and $\partial_{u} \1z \to 0$ as $u\to \pm \infty$ along $\scrip$ in this Bondi frame. There is no explicit assumption on the underlying metric and, at the level of curvature, $\0z$ does not have to become stationary;
\item In full GR, the result follows in a couple of steps if one uses Eqs. (\ref{1dot}) and (\ref{impsi2}) at $\scrip$, which are immediate consequences of Bianchi identities and Einstein's equations at $\scrip$ \cite{np}.
\end{enumerate}
Finally,  note that we do allow kicks in our main analysis;  our assumption \ref{as:psi1} does \emph{not} require that the past and the future rest frame is the same and the ordinary memory does not vanish.

\item The assumption of `past stationarity'  \cite{lbtd,blanchet1}  ---i.e., that the system is stationary before some time $(t=-\tau)$---  \,is extremely strong.  Sometimes one models the behavior of the system in distant past by assuming that it is represented by a binary in a quasi-circular orbit,  rather than by a stationary system. In this case, velocities of the individual bodies would go to zero asymptotically and our two assumptions would still be satisfied. In practice, one may want to avoid the limit $u\to -\infty$ altogether and  simply start at some finite time $u=u_{1}$ sufficiently in the past when the binary is well within the PN regime, and use  (\ref{key2}) in place of (\ref{key1}) with $u_{2} =\infty$. 
\end{enumerate}

\section{Discussion}
\label{s3}

Thanks to the combined technical expertise of the waveform community, the matched filtering procedure based on currently available waveforms has led to the dramatic discovery of several CBCs \cite{LIGOScientific:2018mvr}. The pace of detection has gone up considerably already during the current LIGO-Virgo observation run and,  with a global network of gravitational wave observatories, the detection rate could be as high as 1000 binary black holes coalescences a year, with masses below $100 M_{\odot}$.  The 3G observatories and  LISA will further increase the total rate, have a higher sensitivity and widen the parameter range significantly. These discoveries will enable us to better understand the astrophysics of sources assuming GR, and also to carry out more stringent tests of GR itself.   Furthermore, some experts emphasize that even with the current LIGO-Virgo detectors, more accurate waveforms could lead to the discovery of near or (sub)threshold events and/or events in regions of the parameter space where NR simulations are sparse \cite{ab}. Therefore, it would be extremely useful to have \emph{external} criteria to ensure that the waveforms used in the analysis \emph{do} represent predictions of GR to a very high degree of accuracy.  As we remarked in the Introduction, the conceptual difficulty is that we do not have waveforms from \emph{exact} GR  with which to compare the candidates! Therefore errors are generally estimated by comparing various analytic waveforms  with the NR ones, which are taken to be the `practical substitutes' for the exact waveforms (see e.g.\cite{Pan:2007nw,Lindblom:2008cm,mp-followup,Kumar:2015tha,Kumar:2016dhh,Purrer:2019jcp}).  In Sect.~\ref{s2} we showed that the supermomentum balance laws provide an alternate avenue: One can test the accuracy of a proposed waveform \emph{vis a vis exact GR}  using  Eqs. (\ref{key1}) and  (\ref{key2}). These equations impose an infinite number of constraints on the waveform that must be satisfied in full, non-linear GR.  Since they refer only to the asymptotic structure of space-time, they hold not only for the BH-BH binaries, but also for NS-NS and NS-BH binaries. Violations of these constraints can provide sharp error-bars on any waveform candidate for all CBCs.
Work is in progress to use the constraints to gain further insights in several directions.\vskip0.15cm

The first concerns sub-dominant modes in the (spin-weighted) spherical harmonic decomposition of the waveform. Different models already provide us with several modes beyond the leading (2,2) mode. For example, in the spinning, non-precessing case, there are EOB waveforms that include the (2,2), (2,1), (3,3), (4,4), and (5,5) modes \cite{Cotesta:2018fcv} while in phenom models precessing waveforms that include all $\ell \leq 4$ modes are available (PhenomPv3HM \cite{PhenomPv3HM}). It is natural to ask for the errors one makes in ignoring other modes \cite{Littenberg:2012uj,Brown:2012nn,Capano:2013raa,Varma:2014jxa}. Since  the equality (\ref{smearedkey1}) assumes that $\sigmaz$ on the right side is the exact GR waveform ---which include all modes---  it is can be used to evaluate these errors. Significant violations of this equality could also be used as signals that certain waveforms need further scrutiny. 

In practice it useful to carry out a spherical harmonic decomposition, using $Y_{\ell,m}(\theta, \varphi)$ (for various $\ell, m$) for $f(\theta,\varphi)$ in (\ref{smearedkey1}):
\be \label{smeared2key} \oint \d2vo\, Y_{\ell,m} \Big(GM_{\iz}  -  \f{GM_{i^{+}}}{\gamma^{3} \big(1- \f{\vec{v}}{c} \cdot \h{x}  \big)^{3} }\,\Big) =  \oint \d2vo\,  Y_{\ell,m}(\theta,\varphi) \int_{-\infty}^{\infty}\!\!\rmd u\,  \big[ |\dot\sigmaz|^{2} - \Re(\eth^{2} \dot{\bar\sigma}^{\circ}) \big] (u,\theta,\varphi)\, . \ee
One can use the candidate waveform $\f{1}{2}\big(h_{+}^{\circ} + i h_{\times}^{\circ}\big)$ for $\sigma^{\circ}$ on the right side of this equation, and also in the calculation of the velocity $\vec{v}$ on the left side. Generically, there will be a mismatch between the right and left sides of (\ref{smeared2key}) which will provide an estimate of the error in the candidate waveform for each choice of $\ell, m$. For instance, the $m =0$ modes are usually set to zero in non-precessing models, and they arise in precessing waveforms solely from the time-dependent rotation that takes into account precession. For non-precessing systems, therefore, the $Y_{\ell,0}$ components of the constraint (\ref{smeared2key}) are violated, primarily because the second term on the right side is vanishing by construction. Work in progress shows that for the IMRPhenomD \cite{PhysRevD.93.044007} and SEOBNRv4 \cite{bohe2017improved} models the violation is most significant for $Y_{2,0}$, as one might expect. Errors in different models are essentially the same because their $|\dot\sigmaz|^{2}$ is essentially indistinguishable and the left hand side negligible, being of order $(v/c)^2$. 

However, for other modes the magnitude of the violation of (\ref{smeared2key}) can differ markedly from one class of models to another. This is in particular the case for precessing systems. As a specific example, let us compare SEOBNRv3 \cite{PhysRevD.89.084006,babak2017validating} and IMRPhenomPv2 \cite{PhysRevLett.113.151101} (whose waveforms includes all $\ell = 2$ modes). Consider then a precessing binary system of equal mass black holes, each with spin $0.4$ in the positive $x$-direction (measured at a reference time corresponding to a GW frequency of $0.01$ in units of the total mass). Since these models do not include the (3,2) mode, by fiat the second term on the right hand side of  Eq.~(\ref{smeared2key}) is set to zero in the (3,2) part of this constraint. Therefore again the constraint is violated. However, in this case  the orders of magnitude of the violation are \emph{very} different: while it is $4 \times 10^{-4}$ for SEOBNRv3, it is of order $10^{-14}$ for IMRPhenomPv2,  both in units of $GM_{\iz}$. Therefore ---unlike in the case of the (2,0) mode discussed above--- the discrepancy cannot come solely because the (3,2) mode is ignored; there is another source of systematic error involved. Of course to determine whether these inaccuracies are significant in practice, one would have to take into account  the precise sensitivity of the next generation detectors.

As pointed out in Sect.~\ref{s1}, recently 
these constraints have been used to calculate the probability distribution functions (PDFs) for the total gravitational  memory in various LIGO events \cite{adlkk}. Recall that the the mass and spin of the final black hole serve as inferred observables for LIGO events to date, and comparisons of the posterior PDFs of these observables, obtained from EOB and Phenom waveforms, have generally added to confidence that the accuracy of these waveforms is adequate. The total memory is another inferred  observable associated with  CBCs.  It can also be used to probe systematic errors in given waveforms. 

The strategy can be summarized as follows. Let us set $\ell\ge 2$ and  re-express  (\ref{smeared2key})  by:  (i) carrying out the $u$ integral in the term that is linear in $\dot{\bar\sigma}^{\circ}$ on the right hand side;\,  (ii) using the fact that ${\rm{Im}}\,\eth^{2} \bar{\sigma}^{\circ}$ vanishes at $u=\pm\infty$ because ${\rm{Im}} \2z$ vanishes there;\,  (iii) integrating \,$\eth^{2}$\, by parts and using the identity\,\, $\eth^{2} Y_{\ell,m} = \frac12\sqrt{(l-1)l(l+1)(l+2)}\,\,\, {}_{2}Y_{\ell\,m}$;\,\, and, (iv) moving the result to the left side. We  then obtain\,%
%
%
\be \label{smeared3key} 
\begin{split}
&\frac12\sqrt{(l-1)l(l+1)(l+2)}\;\Big[\lim_{u_{\circ}\to \infty}  - \lim_{u_{\circ}\to -\infty}\Big]  \oint_{u=u_{\circ}} \d2vo\, \,{}_{2}Y_{\ell, m}\, \Re(\bar{\sigma}^{\circ}) \\
&\phantom{\frac12\sqrt{(l-1)}}= GM_{i^{+}} \oint \d2vo\,  \f{Y_{\ell,m}}{\gamma^{3} \big(1- \f{\vec{v}}{c} \cdot \h{x}  \big)^{3} }\, \, +\,\, \oint \d2vo\,  Y_{\ell,m} \int_{-\infty}^{\infty}\!\!\rmd u\, \, |\dot\sigmaz|^{2}  (u,\theta,\varphi)\, .\end{split}\ee
The left hand side of (\ref{smeared3key}) provides the $(\ell,m)$ component of the total gravitational memory. In the framework developed in Sect.~\ref{s2}, $\sigma^{\circ} =0$ at $i^{\circ}$, so the left hand side receives contributions only from $u\to+\infty$, which is  generically non-zero because the past and future Bondi-frames are related by a \emph{supertranslation}  (see, e.g., \cite{np2,aa-rad,aa-yau,gsrw}). This subtlety is important for hybrid waveforms where it is natural to work in the past Bondi frame in the PN regime, thereby setting $\sigma^{\circ}=0$ at $i^{\circ}$, and it is also natural to work with the future Bondi frame in the NR regime, where $\sigma^{\circ}=0$ at $i^{+}$. To obtain a consistent waveform, one then has to incorporate the supertranslation relating the two frames \cite{boyle2016transformations,Adlk2}. However, this term vanishes for all $(\ell, m)$ in the currently available Phenom and EOB waveforms, indicating that they do not incorporate this supertranslation. 

The idea is to \emph{use} our constraint to extract the correct total memory by evaluating the right hand side \cite{garfinkle2016simple,adlkk}. Detailed examination shows that for CBCs that have been observed so far, the first term on the right side (called the linear (or ordinary) memory) is dominated by the second term on right hand side (called the non-linear (or null) memory).  The second term, in turn, is dominated by modes (particularly (2,2)) that are expected to be correctly captured in the available waveforms. Therefore, in various LIGO events one can use the currently available posterior PDFs from EOB and Phenom models to infer the PDFs for total memory. Comparison between these pairs of PDFs for  the $(2,1)$ component of the total memory has signaled \cite{adlkk} a possible source of systematic error that was not revealed by the PDFs of currently used observables.
 
While the total gravitational memory can be evaluated using the above procedure, the calculation of the full (2,0) mode of the waveform \emph{as a function of time} is more challenging. In the NR regime, accurate results have been obtained \cite{pollneyReisswig} using Cauchy-Characteristic Extraction (CCE). It can be seen that  the (2,0) mode has a monotonic behavior, superimposed by some oscillations. This form can be understood conceptually using our constraints: The monotonic part can be directly traced to the (finite time version of  the) non-linear memory term, and the oscillatory, to the (finite time version of  the) linear memory term
in (\ref{key2}).%
\footnote{More precisely, we need to work with a finite time analog of (\ref{smeared3key}), obtained by integrating (\ref{key2}) against $Y_{2,\,0}(\theta,\varphi)$ and setting $u_{1}=-\infty$. In our setup $\sigma^{o}$ vanishes at $u=-\infty$. Therefore, this procedure expresses $\oint \d2vo\, Y_{2,0} (\theta,\varphi)\,\sigma^{\circ}(u_{2},\theta,\varphi)$ as a sum of the (finite-time) non-linear memory that is monotonic, and a (finite-time)  linear memory term which turns out to be oscillatory.}
However, CCE is numerically expensive. Therefore, analytic considerations suggested by the constraints presented in this paper are being used \cite{mikbdLkmpst}  
to fill a gap in the current SXS catalog.

Next, recall that the mathematical GR literature generally assumes that $\sigma^{\circ}$ can be freely specified on $\scrip$. On the other hand we found that it is subject to an infinite number of constraints stemming from the balance laws. How did this come about? The constraints arose because (i) we are considering gravitational waves emitted in CBCs, rather than source-free solutions; and, (ii)  we assumed that the CBCs become stationary in a weak sense both in the distant past and future, in the respective asymptotic rest frames.
Let us therefore reexamine the boundary conditions we imposed in the limits $u \to \pm \infty$ in Sect.~\ref{s2.3.1}. The first has direct physical motivation: We assume that the Bondi news  ---the time derivative of the waveform--- decays sufficiently rapidly so that the total flux of supermomentum across $\scrip$ is well defined. (This is also the necessary and sufficient condition for the flux of Bondi angular momentum across $\scrip$ to be finite. The condition is satisfied in, e.g., the analysis of non-linear stability of Minkowski space given, e.g., in \cite{dcsk}.) This condition guarantees that the system becomes asymptotically stationary  in the weak sense that time derivatives of the components $\4z,\, \3z$ and $\2z$ of the asymptotic Weyl curvature go to zero in any Bondi frame as $u\to \pm\infty$ (if these conditions hold in one Bondi frame, they hold in all). The second condition we imposed is a  strengthening of this notion of asymptotic stationarity: We require that the time derivative of $\1z$ also goes to zero as $u\to \pm \infty$. But now the limits depend on the choice of the Bondi-frame, whence the requirement in the past is imposed \emph{only} in the Bondi frame in which the system is at rest in distant past,  and similarly the requirement in the future is imposed \emph{only} in the Bondi frame in which the system is at rest in distant future.  We saw that, had we demanded the time derivative of $\1z$ vanishes as $u\to \pm\infty$ in the \emph{same} Bondi-frame, the condition would have been physically too restrictive ---in particular it would have ruled out black hole kicks \cite{kicks1,kicks2}! By contrast, our condition requires asymptotic stationarity in a weak sense.  In particular,  in NR simulations  space-time geometry approaches the Kerr  solution in the distant future and PN analysis used in the early phase of coalescence assumes a \emph{much} stronger notion of stationarity in the distant past.  

Finally we would like to clarify a potential confusion about the field $\1z$ on $\scrip$. Global results by Christodoulou and Klainerman \cite{dcsk} on non-linear stability of Minkowski space showed that $\Psi_{4}, \Psi_{3}$ and $\Psi_{2}$  `peel' as $1/r,\, 1/r^{2}$ and $1/r^{3}$ respectively, as implied by the  Newman--Penrose asymptotic conditions \cite{np}. However, in that analysis, $\Psi_{1}$ is guaranteed to fall-off only as $1/r^{7/2}$, rather than as $1/r^{4}$. Therefore, it follows from Eq.~(\ref{weyl}) that  $\1z$ need not exist on $\scrip$ for the class of initial data considered in Ref. \cite{dcsk}. Note that this is not a statement about limits $u\to \pm \infty$ along $\scrip$ considered in this paper; rather, $\1z$ need not exist \emph{anywhere} on $\scrip$. In particular, this means that the angular momentum of the system would be ill-defined at \emph{any} retarded time $u=u_{o}$ \cite{td,ak-thesis}. From a physical viewpoint, therefore, this analysis caters to too broad a class of systems. In CBC, in particular, angular momentum \emph{is} well-defined and constitutes an important parameter in the system characterization. Indeed, even in the vacuum case to which Ref. \cite{dcsk} restricts itself to, Chrusciel and Delay \cite{pced} have shown that there is a non-linear neighborhood of Minkowski initial data that evolves to a unique global solution in which the Newman--Penrose peeling holds, i.e. $\1z$ has a well-defined $C^{k}$ limit to $\scrip$. Finally, note that all these global results refer to vacuum (or electrovac) situations in which there is outgoing \emph{as well as incoming} gravitational radiation. Physically we are much more interested in gravitational waves produced by sources ---such as compact binaries--- with no incoming radiation on $\scrim$. Among solutions covered by the currently available global existence theorems, the only one with no incoming radiation is Minkowski space! Therefore, these theorems do not offer direct guidance for what the appropriate fall-off conditions would be for systems undergoing compact binary coalescence.

To summarize, our analysis is based on results that have been well-known in the mathematical relativity community for many years. The new feature is the recognition that (a weaker form of) the assumption of asymptotic stationarity used in the waveform community, together with the knowledge of the total mass of the system, leads to an infinite set of constraints on the CBC waveforms in general relativity.  Thus, the analysis provides a new avenue to evaluate the accuracy of candidate waveforms vis a vis exact general relativity. This tool has already been used to probe the accuracy of waveforms and differences in the underlying physics of the currently used waveform models in \cite{adlkk}, and  to improve them  in \cite{mikbdLkmpst}. These examples can be considered as a proof of principle that the tool will be even more useful as the statistical errors decrease with new generations of gravitational wave detectors. 

\section*{Acknowledgments}
 We would like to thank K. G. Arun, A. Gupta, and B. Sathyaprakash for comments on the manuscript; Bala Iyer and Luc Blanchet for extensive correspondence on the PN methods;  Alessandra Buonanno for discussions on EOB, and Larry Kidder, on NR; and Badri Krishnan for several detailed discussions on waveforms in general and Phenom models in particular.  We are also indebted to participants of the APS meeting in Denver, IGC$@$25 conference at Penn State,  GR22/Amaldi13 conference in Valencia, and the Discussion meeting on Future of Gravitational Waves at ICTS, Bangalore for questions and suggestions. This work was supported in part by the NSF grant PHY-1806356, grant UN2017-92945 from the Urania Stott Fund of Pittsburgh Foundation and the Eberly research funds of Penn State.

\begin{appendix}
\normalsize
  
\section{Null infinity and emergence of supertranslations}
\label{a1}

In Sect.~\ref{s2}  we used several known facts about the structure of null infinity and properties of fields thereon. Since some in the PN and NR communities may not be familiar with them, in this Appendix we present a brief summary, focusing only on those features that we need. 

Eventhough Einstein, Eddington and others explored the properties of gravitational waves in the weak field approximation around Minkowski space soon after the discovery of general relativity, there was considerable confusion about the reality of gravitational waves in \emph{full, nonlinear} GR for several subsequent decades largely because of the coordinate freedom: What appeared to be a wave-like behavior in one coordinate system could appear stationary in another.  This confusion was resolved only in the 1960s when Bondi, Sachs and others showed that one can unambiguously disentangle gravitational waves by moving away from isolated sources in retarded \emph{null directions}, i.e., in the usual terminology, by taking the limit $r \to \infty$ keeping the retarded time constant $u={\rm const}$. The asymptotic boundary conditions introduced by Bondi and Sachs  \cite{bondi,sachs1} were geometrized by Penrose \cite{rp}  through the notion of a conformal completion of spacetime, i.e., by attaching to spacetime a 3-dimensional  boundary $\scrip$, representing \emph{`future null infinity'}.

These frameworks provided a definitive, coordinate invariant characterization of gravitational radiation in asymptotically flat spacetimes and introduced techniques to analyze its properties in exact, non-linear general relativity. However, initially there were concerns as to whether the underlying assumptions are too strong to be satisfied by realistic isolated systems such as compact binaries (see, e.g., \cite{lbtd}). The current consensus is that they are not too strong. In particular, the asymptotic form of the PN metric is completely consistent with the Bondi--Sachs--Penrose framework, as shown for instance by Theorem 4 in \cite{blanchet1}. Similarly,  the key  notions of this framework ---such as the radiation field $\4z$, the Bondi news $N$, and the asymptotic shear $\sigmaz$--- and their properties are heavily used in numerical simulations of  waveforms and calculations of energy and momentum flux in NR. These notions and properties are summarized in Sect.~\ref{s2}.

 The detailed analysis of gravitational radiation at null infinity brought to forefront an unforeseen result that plays a key role in Sect.~\ref{s2}: Even though spacetimes representing isolated gravitating systems are asymptotically Minkowskian, the asymptotic symmetry group is \emph{not} the Poincar\'e group $\P$, but rather an infinite dimensional generalization thereof, the BMS group $\B$. This is a consequence of the fact that gravitational radiation is on a genuinely different footing ---from, say, the electromagnetic one--- in one important respect. It  introduces ripples in spacetime curvature that extend all the way to infinity, i.e., to $\scrip$, making it impossible to single out a \emph{preferred} Poincar\'e group $\P$ using asymptotic Killing vectors.  This difficulty can be seen in concrete terms as follows.  Suppose we have a metric $g_{ab}$ that is asymptotically flat in the sense of Bondi and Sachs; so
 it approaches a Minkowski metric $\eta_{ab}$,  with $g_{ab} = \eta_{ab} +\mathcal{O}(1/r)$,  as $r \to \infty$ keeping $u=t-r$ constant.  Therefore, Poincar\'e  transformations of $\eta_{ab}$  provide us with asymptotic Killing fields for $g_{ab}$. Now consider  a diffeomorphism $ t \to t^{\prime} = t +f(\theta,\varphi), \,\, \vec{x} \to \vec{x}^{\prime} = \vec{x}$ where $t,\vec{x}$ are Cartesian coordinates of $\eta_{ab}$.  This is an \emph{angle dependent  translation}, whence the metric  $\eta_{ab}$ is sent to a \emph{distinct} flat metric $\eta^{\prime}_{ab}$.%
 \footnote{We chose a time-translation just for definiteness: the argument continues to hold if the `angle dependent translation'  is generic.}
 One can verify that since $g_{ab}$ approaches $\eta_{ab}$ as $1/r$ \`a la Bondi--Sachs, it also approaches $\eta^{\prime}_{ab}$ as $1/r^{\prime}$ \`a la Bondi--Sachs!  Therefore, the Poincar\'e transformations of $\eta^{\prime}_{ab}$ are also asymptotic Killing fields of our physical $g_{ab}$. But since the two Minkowski metrics are distinct, their isometry groups $\P$ and $\P'$ are also distinct.  The BMS group can be interpreted as a `consistent union' of Poincar\'e groups  associated with all these Minkowski metrics, related to one another by `angle dependent translations.'  These are known as \emph{supertranslations}. Detailed examination  brought out another subtlety: All these Poincar\'e groups define the same translation subgroup asymptotically, whence the BMS group does admit a canonical, 4-dimensional translation subgroup $\T$  \cite{sachs2}.  However, the Lorentz subgroups $\Lor$ of various Poincar\'e groups are different even asymptotically. Recall that the Poincar\'e group $\P$ admits a 4-parameter family of Lorentz subgroups ---each of which defines rotations and boosts about one specific origin in Minkowski space---   related to one another by a translation. By contrast,  the BMS group $\B$ admits an infinite parameter family of Lorenz subgroups that are related to one another by supertranslations.  This gives rise to the well-known  `supertranslation ambiguity' in the notion of angular momentum at null infinity. We discuss this issue in detail in \cite{Adlk2}, again in the context of CBC.
 
 In this paper we focused on supertranslations.  Just as the translational symmetries of  the Minkowski metric lead to the notion of energy-momentum for fields in Minkowskian physics,  supertranslation symmetries on $\scrip$ lead to the notion of \emph{supermomenta}. In the case of energy-momentum, we have two different quantities available at $\scrip$. The first is the Bondi 4-momentum  ---a 2-sphere integral on a cross-section $C$ at $\scrip$, representing the energy momentum of the system, left over at the retarded instant of time $u=u_{0}$ defined by $C$. The second is the notion of flux of energy-momentum carried away by gravitational waves through a `patch'  $\Delta {\scrip}$ of $\scrip$. As a consequence we have a balance law: The difference between the Bondi 4-momentum evaluated on two different cross-sections $C_{1}$ and $C_{2}$ of $\scrip$ is the flux of energy-momentum across the patch $\Delta\scrip$ bounded by them. It turns out that the same is true for supermomentum.  Thus we have an infinite number of balance laws ---Eqs.~\eqref{bal2}  (or ~\eqref{key2}) in the main text--- each characterized by a function on a 2-sphere defining the supertranslation. As we discussed in Sect.~\ref{s2}, these lead to an infinite set of constraints ---imposed by full, non-linear GR--- that any waveform must satisfy in a CBC. 

\end{appendix} 


\bibliography{references}

\end{document}